\documentclass[%
 aip,
 amsmath,amssymb,
 reprint,%
]{revtex4-1}

\usepackage{graphicx}% Include figure files
\usepackage{dcolumn}% Align table columns on decimal point
\usepackage{bm}% bold math
\usepackage[utf8]{inputenc}
\usepackage[T1]{fontenc}
\usepackage{mathptmx}
\usepackage{empheq}
\usepackage{subfigure}
\usepackage{xcolor}
\usepackage{scalerel,amssymb}
\usepackage{empheq}
\usepackage[colorlinks]{hyperref}
\hypersetup{
colorlinks=blue,
linkcolor=blue,
urlcolor=blue,
filecolor=blue,
citecolor=blue
}

\usepackage{array}

\usepackage{tabularx}
\usepackage{hhline}

\usepackage{gensymb}
\usepackage{tikz}             %for drawing circles
\usepackage{textcomp, gensymb} %for rangle symbols
\begin{document}

\preprint{AIP/123-QED}

%\title{How to quantify hydrodynamic effects in self-diffusiophoretic systems}
%\title{Self-diffusiophoresis in hydrodynamic and equivalent non-hydrodynamic models}
%\title{A method to quantify hydrodynamic effects in self-diffusiophoretic systems: application to an active chemoattractive polymer}
%\title{Self-diffusiophoretic Brownian dynamics: characterization of hydrodynamic effects for an active chemoattractive polymer}
\title{Diffusiophoretic Brownian dynamics: characterization of hydrodynamic effects for an active chemoattractive polymer}

\author{Surabhi Jaiswal}
\email{surabhi19@iiserb.ac.in}
\affiliation{Department of Physics, Indian Institute of Science Education and Research Bhopal, Bhopal 462066, India}
\author{Marisol Ripoll}
\email{m.ripoll@fz-juelich.de}
\affiliation{Theoretical Physics of Living Matter, Institute for Advanced Simulation, 
Forschungszentrum J\"ulich, 52425 J\"ulich, Germany}
\author{Snigdha Thakur}
\email{sthakur@iiserb.ac.in}
\affiliation{Department of Physics, Indian Institute of Science Education and Research Bhopal, Bhopal 462066, India}

% \author{Surabhi Jaiswal\textit{$^{1}$}, Marisol Ripoll\textit{$^{2}$} and Snigdha Thakur\textit{$^{1}$}  }
% \email{sthakur@iiserb.ac.in}
%  \affiliation{\textit{$^{1}$}Department of Physics, Indian Institute of Science Education and Research Bhopal, Madhya Pradesh, 462066, India\\ \textit{$^{2}$}Institut f\"{u}r Festk\"{o}rperforschung, Forschungszentrum J\"{u}lich, D-52425 J\"{u}lich, Germany}

\date{\today}% It is always \today, today,

\begin{abstract}
The phoretic Brownian dynamics method is shown here to be an effective approach to simulate the properties of colloidal chemophoretic based systems. 
The method is then optimized to allow for the comparison with 
results from multiparticle collision dynamics, a hydrodynamic method with explicit solvent, which can also be employed in the case of chemoattractive polymers.
In order to obtain a good match of the conformational equilibrium properties of the models without and with explicit solvent, we propose a modified version of the phoretic Brownian dynamics accounting for the explicit solvent induced swelling.
In the presence of activity, chemoattractive polymers show a transition to a compact globular state and hydrodynamics have a non-trivial influence in the polymer collapse times.
The phoretic Brownian method can then be applied to much longer polymers, which allows the observation of a non-monotonous growth of both, the radius of gyration and the relaxation time with polymer length, for such chemoattractive active polymers. 
\end{abstract}

\maketitle

%\tableofcontents

\section{\label{sec:introduction}Introduction}

Diffusiophoresis refers to the thrust that colloidal particles experience when surrounded by a solvent with an intrinsic concentration gradient~\cite{anderson1989,sen2009microbots,sen2011motor}. 
For colloidal particles with an asymmetric chemical reaction catalyzed on the particle surface, the phoretic thrust results in the colloid directed motion, which is usually referred to as self-diffusiophoresis, or chemical activity~\cite{kapral2007nanodimers,kapral2013jcp,snigdha2012collective,robertson2018synthetic,kapral2008jcp,shim2022diffusio,reigh2018diffusiophoretically,meredith2022self-propelleddrop}. Other than chemical gradient, activity has also been similarly introduced via thermal~\cite{jiang2010janus,yang2011nanoswimmers,Burelbach2017colloidal}, magnetic~\cite{suwa2011magnetoanalysis,benelmekki2012magnetophoresis,peyer2013microrobots}, or electric~\cite{hatlo2011translocation,sabass2012selfpropelled,zhan2019enhanced} field gradients, which are referred to as thermo-, magneto-, or electrophoretic active particles. 
These active particles drive therefore the system out of equilibrium, becoming the synthetic equivalent of biological active matter, for which there is a broad number of examples such as insects, bacteria, or molecular motors
~\cite{Bechinger2016activeparticles,hu2016entangled,MENZEL2015tuned,Magistris2015introduction,Gompper2020motile,sramaswamy2010review}. 
Very different from their passive counterparts, active matter systems have largely shown to exhibit a prolific behavior, with properties that strongly depend on the system dimensionality, details of structure of the active particles, specific interactions, or the presence of the hydrodynamic interactions~\cite{ elgeti2015physics,wagner2021collective,palacci2013living,pohl2014dynamic,theurkauff2012dynamicclust,Huang2017chemotactic,stenhammar2014phase,wysocki2014cooperative,delfau2016collective}.

Polymeric systems, where chemical activity regulates structure and dynamics are of particular interest. Bio-polymers like DNA, chromatin, or actin have a functionality provided by the chemical activity of motors/enzymes associated to them~\cite{narlikar2013mechanisms,chaffey2003}. 
In chromatin, the correlated motion of active units on chromosomes gives rise to spontaneous segregation and domain formation~\cite{ganai2014chromosome,andriy2023genomeorganization}.
Motivated by these systems, there is a recent increasing interest on the design of synthetic active polymers and on the understanding of their behavior both theoretically~\cite{ghosh2014semiflexible,eisenstecken2016polym,eisenstecken2017polactivenoise,lisa2023ringpol}, and experimentally~\cite{biswas2017catalystcoated,biswas2021icetemplating,shafiei2020activebath}.
These filaments have shown to exhibit very diverse phenomena like swelling~\cite{eisenstecken2016polym,anand2020flexiblepol,das2021pol,kaiser2015flexchain}, coil-to-globule transition~\cite{bianco2018activepol,bianco2021activering,sandeep2023activering,duman2018collective}, or spiral formation~\cite{sarkar2017spontaneous,chelakkot2014flagellardynamics}. 
For the specific case of phoretically induced activity,  a chain of chemically active colloids has demonstrated similar coil-to-globule transition with a decreasing Flory exponent with activity in a limited range of activities and sizes~\cite{namita2022collapse}.  
 
Computer simulations play a crucial role in the understanding of the active matter systems~\cite{kremer2000computer,shaebani2020computational,Gompper2009,kapral2008book}. In the context of self-phoresis, Janus-like particles have been investigated with different simulation approaches~\cite{reigh2018diffusiophoretically,snigdha2010bound,fedosov2015thermophoretic,roca2022collective,hartmut2020self-prop,shen2018hydrodynamic,hossain2022janusrods}.
In some of these techniques, the phoretic self-propulsion is directly accounted for by a constant force considered in addition to the Brownian colloidal particles, which does therefore not include any hydrodynamic interactions~\cite{golestanian2012collective,pohl2014dynamic,liao2020self-assembly,telezki2020activeparticles}. Other approaches consider both phoresis and hydrodynamic interactions via an explicit solvent, with an intrinsic inhomogeneity~\cite{kapral2007nanodimers,andreas2023hydrodynamics,shen2018hydrodynamic}. 
To disregard the effect of hydrodynamics is not always properly justified, although to quantify their effect is only possible via a comparison of the same systems with two approaches that differ as little as possible besides the consideration of hydrodynamics.
To this effect, there are two standard approaches, those employing Brownian Dynamics, to which an implicit solvent can be additionally considered~\cite{Pham2008Browniandynamics}, or those considering an explicit solvent, whose effect can be eventually disregarded~\cite{kikuchi2002polymer,ripoll2007epje}. The application of these approaches is not yet possible to cases where phoresis is relevant.

In this work, we first adapt the phoretic Brownian dynamics model (Ph-BD), which has been recently developed for thermophoretic self-propelled Janus dimers~\cite{roca2022self,roca2022collective}, to consider the details of diffusiophoresis. 
Then, we apply the method to a dimeric structure and compare the resulting self-propulsion velocity and rotational diffusion to the same quantities measured in simulations with the multiparticle collision dynamics method (MPCD), where hydrodynamic interactions are considered. 
This comparison allows for a precise parameter mapping of the two methods, which is necessary to provide a reliable quantification of the effect of hydrodynamics in diffusiophoretic systems. 
We use both methods to simulate a passive polymer, where the explicit solvent shows to account for a repulsion between beads which is not present in the Brownian approach. 
An additional soft repulsion term is then considered into the repulsive phoretic Brownian dynamics model (RPh-BD). 
The application of the three approaches to the simulation of a chemoattractive polymer, shows that the RPh-BD approach is able to capture most of the conformational characteristics brought by the hydrodynamic model, but that the influence in the dynamic properties is more subtle. 
Finally, we also employ both Ph-BD and RPh-BD methods to investigate the properties of chemoattractive polymers with larger number of monomers than those simulated in the presence of hydrodynamics.   

\section{Simulation models}
\label{sec:model}

\subsection{Phoretic Brownian Dynamics {(Ph-BD)}}
\label{sec:phbd}

Self-diffusiophoresis has been studied in the context of Janus particles which are composed by two different surface properties~\cite{walther_2013}. Numerous works have focused in the investigation of dimeric Janus colloids, where each bead has different composition~\cite{kapral2008jcp,wysocki2014janus}. Typically one of the beads acts as a cataliser, inducing a chemical reaction on its surface, which creates a gradient in the concentration of product. The second bead has a surface that phoretically react to the concentration gradient by inducing a thrust towards the catalytic source (chemoattractive), or against it (chemorepulsive). 

The phoretic model we propose here follows closely the one  proposed for self-thermophoretic dimeric colloids~\cite{roca2022self}, where only the details of the phoretic force need to be slightly adapted. For the sake of completeness, we specify here a general case where $N$ beads are arranged in various dimers or multimeric structures. 
The time evolution of the beads is given by overdamped Brownian Dynamics,  which describes particle motion at low Reynolds number conditions~\cite{roca2022self,pathria2016statistical},
\begin{equation}\label{eq:bd}
    \dot {\bf r}_{i}(t) = \frac{{\bf F}_{i}({\bf r}_{i}(t))}{\mu_i} + \sqrt{\frac{2k_{B}T}{\mu_i}} \xi_{i}(t).
\end{equation}
where ${\bf F}_{i}$ is the total force acting on bead $i$ due to all other beads and depends on the bead type, $k_{B}$ is the Boltzmann constant, and $T$ the average solvent temperature. The random term $\xi_i$ is due to the fluctuations of the implicit solvent, has zero mean $\langle \xi_{i} \rangle=0$, and a delta-correlated Gaussian relation $\langle \xi_{ip}\cdot\xi^{jq}\rangle=\delta(t-t^{\prime})\delta_{ij}\delta_{pq}$, with $j=1,....,N$, and $p,q=x,y,z$. 
The friction coefficient $\mu_i$ is considered to accomplish the Stokes-Einstein relation as $\mu_i=C_{f}\pi\eta R_i$, with $\eta$ the solvent viscosity, $R_i$ the bead radius, and $C_{f}$ a numerical factor which depends on the colloid surface boundary conditions (stick $C_{f}=6$ or slip $C_{f}=4$)~\cite{roca2022self, hansen2013theory}. 

The total force on each particle depends on the particle functionality, such that we distinguish in general three types of particles (source, linkers, and phoretic), with two types of forces. 
Source~(S) particles model beads with catalytic surface function such that they are the origin of concentration gradient, without responding to it.  Phoretic~(P) particles model beads whose surface has a significant phoretic response to a solvent chemical gradient. Linkers~(L) particles model beads which do not generate any chemical gradient, and have also a negligible response to any external gradient.   

All monomer types are affected by the force ${\bf F}^{A}$, which is the sum of two contributions
\begin{equation}\label{eq:sourceforce}
    {\bf F}^{A}_{i}({\bf r}_{i}) = {\bf F}_{H,i}({\bf r}_{i}) + {\bf F}_{EV,i}({\bf r}_{i}).
\end{equation}
Consecutive beads of the same structure are connected by a strong harmonic force ${\bf F}_{H,i}$ given by the harmonic potential 
\begin{equation}\label{eq:harmonic}
U_{H}(r_{ij}) = \frac{\kappa}{2}(r_{ij}-r_0)^2,
\end{equation}
where $r_{ij}=|{\bf r}_{i}-{\bf r}_{j}|$ is the beads separation, $r_0$ the internuclear separation, and $\kappa$ is the spring constant.
Furthermore, non-consecutive beads have an excluded volume interaction force ${\bf F}_{EV,i}$ given  by the Weeks-Chandler-Andersen (WCA) potential
\begin{equation}\label{eq:nonneighpotential}
%\begin{split}
U_{WCA}(r_{ij}) = 4\epsilon \Bigg[\Big(\frac{\sigma_{ij}}{r_{ij}}\Big)^{12}-\Big(\frac{\sigma_{ij}}{r_{ij}}\Big)^{6} + \frac{1}{4}\Bigg] \Theta(r_{ij}- r_{c}), 
%~~~~~0<r_{ij}\leq r_{c} \\  & = 0, ~~~~~\text{otherwise}.
%\end{split}
\end{equation}
where $\sigma_{ij}=R_{i}+R_{j}$ corresponds to the sum of the particles radius, $\epsilon$ to the potential steepness, $\Theta$ to the Heaviside function, and $r_c=2^{1/6}\sigma_{ij}$ to the potential cutoff radius. 

For source and linker type of beads, ${\bf F}^{A}$ is already the total applied force. However,
the force on the phoretic beads requires the consideration of one additional term to model the effect of a diffusiophoretic or chemophoretic force, ${\bf F}_{c}$
\begin{equation}\label{eq:phforce}
    {\bf F}^P_{i}({\bf r}_{i}) = {\bf F}^{A}_{i}({\bf r}_{i}) + {\bf F}_{c,i}({\bf r}_{i}). 
\end{equation}
The chemophoretic force is proportional to the chemical gradient and the characteristic particle properties, such that we consider
\begin{equation}\label{eq:chemforce}
    \mathbf{F}_{c,i}({\bf r}_{i}) = -\alpha_{c}\frac{k_{B}T}{\rho}{\bf\nabla} c({\bf r}_{i}).
\end{equation}
\noindent
Here ${\bf\nabla} c({\bf r}_{i})$ is the effective chemical gradient and $\rho$ is the average total density. 
The chemophoretic diffusion constant $\alpha_{c}$  refers to the particle phoretic properties, and is defined in Eq.~(\ref{eq:chemforce}) as a dimensionless number. %%
The magnitude of $\alpha_{c}$ describes the strength of the particle surface phoretic effect, and the sign is positive for chemo-repellent particles and negative for chemo-attractive ones. The chemical gradient ${\bf\nabla} c({\bf r}_{i})$ considers the effect experienced by a phoretic particle $i$ due to all chemical sources in the system which, similarly to the case of the thermophoretic force~\cite{roca2022self,roca2022collective},  can be approximated by 
\begin{equation}\label{eq:chemgrad}
   {\bf\nabla} c({\bf r}_{i})  = \sum_{j}\frac{ \Delta c }{(r_{ij}+R_{p})(r_{ij}-R_{p})} R_{s}{\hat {\bf r}}_{ij}, % {\mathbf{r}}_{ij},
\end{equation}
where $\Delta c$ is the difference of concentration at the surface of the source bead $c(R_{s})$ and the bulk $c(r\rightarrow\infty)$. This difference is an input parameter for the simulation modelling, both the strength of the catalytic reaction and the amount of reactant. The summation in Eq.~(\ref{eq:chemgrad}) accounts for  $j=1,\ldots N_s$ with $N_s$ the number of source (S) beads, which considers the effect of all the sources to be simply additive. 

The strength of the phoretic effect is determined by the phoretic force and therefore by both $\Delta c$ and $\alpha_c$. However, for the investigation of self-propelled systems it is standard, and meaningful, to characterize the activity in terms of a dimensionless P\'{e}clet number ($Pe$) which relates advection and diffusion, such as 
\begin{equation}\label{eq:pec}
    Pe = \frac{V_{s}}{D_{r}R_p},
\end{equation}
where $V_{s}$ is the average swimmer self-propulsion velocity, $D_{r}$ the rotational diffusion coefficient and $R_p$ is a relevant system dimension, which here we take as the phoretic particle radius. With this expression, the simulations results of the phoretic Brownian model  can be compared with results of other simulation methods or experiments.  
With the phoretic Brownian dynamics, the values of $V_s$ and $D_r$ can be explicitly measured in simulations, or evaluated from the model parameters. 
The rotational diffusion coefficient can be estimated from the system parameters as 
\begin{equation}\label{eq:dr}
D_r= \frac{k_BT}{8\pi\eta R^3},
\end{equation}
with $R$ the swimmer radius. The average self-propulsion velocity is related to the phoretic force by $F_c=-\mu V_s$,  with $\mu$ the average value of the particle friction coefficients $\mu_i$, such that 
\begin{equation}\label{eq:alphac}
    \alpha_{c} = \frac{\mu V_{s}\rho}{k_{B}T\langle\nabla c\rangle}. 
\end{equation}   

\subsection{Repulsive Phoretic Brownian Dynamics {(RPh-BD)}}
\label{sec:rphbd}

For later reference, we also introduce a variant of the Ph-BD method which takes into account an additional repulsive force ${\bf F}_{R}$ among all monomers. 
%{\color{blue} The origin of this force will be discussed later.}
We consider a finite-range soft repulsive force similar to those employed in DPD~\cite{Hoogerbrugge1992dpd,soleymani2020dpd}
\begin{equation}\label{eq:frep}
    \mathbf{F}_{R,i}({\bf r}_{i}) = \sum_j a_{ij} \Bigg[1 -\frac{r_{ij}}{r^{\prime}_c}\Bigg] {\hat {\bf r}}_{ij} \Theta(r_{ij}- r^{\prime}_{c}). 
\end{equation}
Here the strength of the repulsion $a_{ij}$ and the range of the interaction $r^{\prime}_c$ are parameters that can be varied according to the model requirements.

\subsection{Multi-particle Collision Dynamics} 
\label{sec:mpc}

In the presence of hydrodynamic interactions, an explicit solvent method has been used to simulate chemophoretic self-propelled dimeric and polymeric particles \cite{kapral2007nanodimers,snigdha2010bound,colberg2017jcp}, which is completely different from the Brownian approach described above.
This is a hybrid molecular dynamics (MD), multi-particle collision dynamics (MPCD) approach, where  the dimeric dynamics and the interaction between dimer and solvent is through MD, while for solvent-solvent interactions are given by MPCD. The dimer dynamics considers simply a MD scheme, with the force given in Eq.~(\ref{eq:sourceforce}) with the harmonic and excluded volume interactions in Eq.~(\ref{eq:harmonic}) and Eq.~(\ref{eq:nonneighpotential}). 
The dimer is then immersed in the MPCD solvent. 

The MPCD solvent particles have mass $m$ and perform alternating streaming and collision steps. In the streaming step, the solvent particles move ballistically for a certain time $\tau_c$. After this time, the collision step is performed, wherein the solvent particles are sorted into the collision cubic cells of size $a$ with an applied grid shift to restore the Galilean invariance~\cite{ihle_2001}. Each solvent particle interact with all the other within the same collision cell via %%
\begin{equation}\label{eq:velrot}
    { {\bf v}}_{k }(t+\tau_{c}) = {{\bf v}}_{cm, k}(t) + {\bf \Re}(\phi) ({\bf {v}}_{k}(t) - {{\bf v}}_{cm, k}(t)).
\end{equation}
Here, ${\bf v}_{cm, k}(t)$ is the center of mass velocity of the collision cell where particle $k$ is sorted, and ${\bf \Re}(\phi)$ is the rotation matrix with angle $\phi$, around an axis randomly chosen in each cell. 
In this way, particles interchange locally momentum, but conserve mass, collision cell linear momentum and kinetic energy. This approach has repeatedly shown to properly include the effect of a hydrodynamic solvent~\cite{Gompper2009,kapral2008book}.

In order to consider the chemical character of the solvent when interacting with the colloids surfaces, two solvent particle types are distinguished, $N_A$ particles of type $A$ and $N_B$ particles of type $B$. 
When particles of type $A$ come closer than $2^{1/6}R_s$ of the surface of a catalytic bead (where $R_s$ is the source bead radius)  chemical reaction, $A+S \rightarrow B+S$ takes place~\cite{kapral2007nanodimers,kapral2008jcp}. 
This reaction on $S$ generates a non-uniform concentration of product ($B$) around  the phoretic particle ($P$). To mantain a constant ratio of particles $A$ and $B$ in the overall system, all the $B$ particles are converted back to $A$ when they diffuse sufficiently far away from the source particle. %%
On the other hand, the interaction of the solvent particles with the dimer beads occurs via a WCA potential, as in Eq.~(\ref{eq:nonneighpotential}). For such interaction, only the colloid radius is considered $\sigma_{ij}=R_i$, and the energetic interchange varies with the nature of the particles, namely solvent $A$ or $B$ with beads $S$ or $P$. The choice made here is that all interactions are given by $\epsilon$, and only the interaction of the product particles $B$ with the phoretic beads $P$ is different, $\epsilon_B$~\cite{kapral2007nanodimers,kapral2008jcp}. This difference in force along with the non-uniform solvent gradient around $P$ due to the chemical reaction, leads to an effective phoretic thrust.
The strength and direction of such phoretic force is determined by the ratio between $\epsilon_B$ and $\epsilon$, and the precise determination of the P{\'e}clet number for a self-propelled structure requires the determination of $V_s$ and $D_r$ in Eq.~(\ref{eq:pec}).

\subsection{Mapping of HI- and HI+ schemes for self-phoretic dimers: simulation parameters choice}  
\label{sec:mapping}

The two methods presented in the previous sections are in principle completely independent, such that the relevant parameters can be chosen attending separate requirements. 
Here, we are however interested in quantifying the effect of hydrodynamic interactions, such that we intend to compare systems that differ as little as possible, besides the existence of HI, for which we need to optimize the choice of parameters. The mapping is made considering a single self-propelled dimer composed of one source bead and one phoretic bead and performing simulations in the HI+ case, this is with the MD-MPCD method and the HI- case, this is with the Ph-BD method. Note that later we will also use the case where Ph-BD also includes repulsive interactions, which we will refer to as HI-R.

The MPCD simulations quantities are given in units of the particle mass, the system thermal energy, and the cell length, this is $m=k_{B}T=a=1$. The average number of solvent particles in a collision cell is chosen to be $\rho=10$, the rotation angle $\phi=\pi/2$, and the collision time $\tau_c=0.5$. With these parameters the solvent viscosity can be calculated to be $\eta=3.1$~\cite{ihle2003,ripoll2005,tuzel2006}. The radius of the beads, unless otherwise specified, are $R_{s}=R_{p}=1$, the harmonic spring constant $\kappa=30$, the internuclear separation $r_{0}=R_s+R_p+0.45$, and the MD integration with the velocity Verlet algorithm, with time step $\Delta t=0.01$. The energy parameter is $\epsilon=1$, and $\epsilon_{B}$ is modified in order to vary $Pe$.

\begin{figure}[ht!]
     \includegraphics[width=0.4\textwidth]{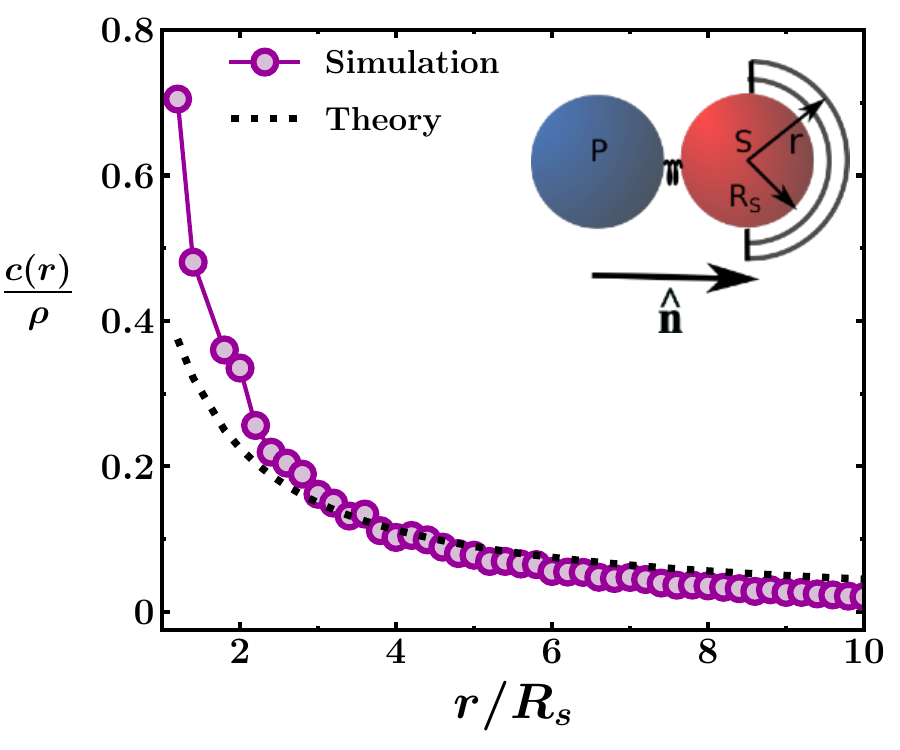}
     \caption{Variation of the normalized product molecule concentration, $c(r)/\rho$, as a function of $r$, the distance from the surface of the source bead, $S$. 
Symbols correspond to the averages from simulations and the dotted line to the $\sim r^{-1}$ prediction from the Laplace equation.
     The inset shows the schematic for the dimer and the hemisphere chosen to calculate $c(r)$, and the large arrow indicates the direction of phoretic force.}
     \label{fig:gradient}
 \end{figure}

The average self-propulsion velocity of the dimer in MPCD is determined from the simulations as  $V_{s}=\langle{\bf V}_{cm}\cdot {\bf\hat{n}}\rangle$, with ${\bf V}_{cm}$ the dimer center of mass velocity, and $\hat{\bf n}$ the unit vector from $P$ to $S$ (as shown in the inset of Fig.~\ref{fig:gradient}). The rotational diffusion coefficient $D_{r}$ is computed by the long time behavior of the mean squared angular displacement $\Delta \theta^2(t) = \langle (\theta(t) - \theta(t_0))^2\rangle$. With the above choices of parameters and $\epsilon_B=0.01$ we obtain $V_{s}=0.03$, $D_{r}=0.0072$, and hence $Pe=4.2$ for the HI+ case. %%
In order to gain further insight, we also compute the concentration of product in concentric half shells around the source bead as indicated in the sketch in Fig.~\ref{fig:gradient}.   
The profile in Fig.~\ref{fig:gradient} displays the concentration of product around the source bead, which shows that the phoretic bead is exposed to a gradient in composition. The decay of concentration of product in Fig.~\ref{fig:gradient} nicely agrees with $\sim r^{-1}$ decay, this is with the inverse of the distance from the bulk average density to a vanishing value, as expected from the Laplace equation, and is consistence with Eq.~(\ref{eq:chemgrad}) for the concentration gradient. The value closest to the source surface depends on the thickness of the measuring shells, and the overall pre-factor is influenced also by the presence of the phoretic bead, as well as finite size effects~\cite{shangyik2015}.

The Ph-BD simulation quantities are given in units of particle mass, and thermal energy, $m=k_BT=1$. The bead radius $R_s$ could be considered as a length unit, but this needs to be let open due to noise intensity regulation, as required for the mapping. First we consider $\rho=10$, $\eta=3.1$ which are the direct inputs from the MPCD model, and $R_p=R_s$. The integration is performed with the Euler algorithm with time step $\delta t = 0.001$. 
For the friction coefficient $\mu=C_{f}\pi\eta(R_{s}+R_{p})$ is considered,  and we take $C_{f}=3$ in order to match the behavior found for MPCD without angular momentum conservation~\cite{yang2015ang}. 
Then we need to find the values that controls i)~the intensity of the phoretic force in Eq.~(\ref{eq:chemforce}), which we tune to fit the same self-propelled dimer velocity as in the HI+ case, and ii)~the rotational diffusion in Eq.~(\ref{eq:dr}), which we tune to match then both $D_r$ and $Pe$ of the MPCD model. 

\begin{figure}[ht]
     \centering
     \includegraphics[width=0.4\textwidth]{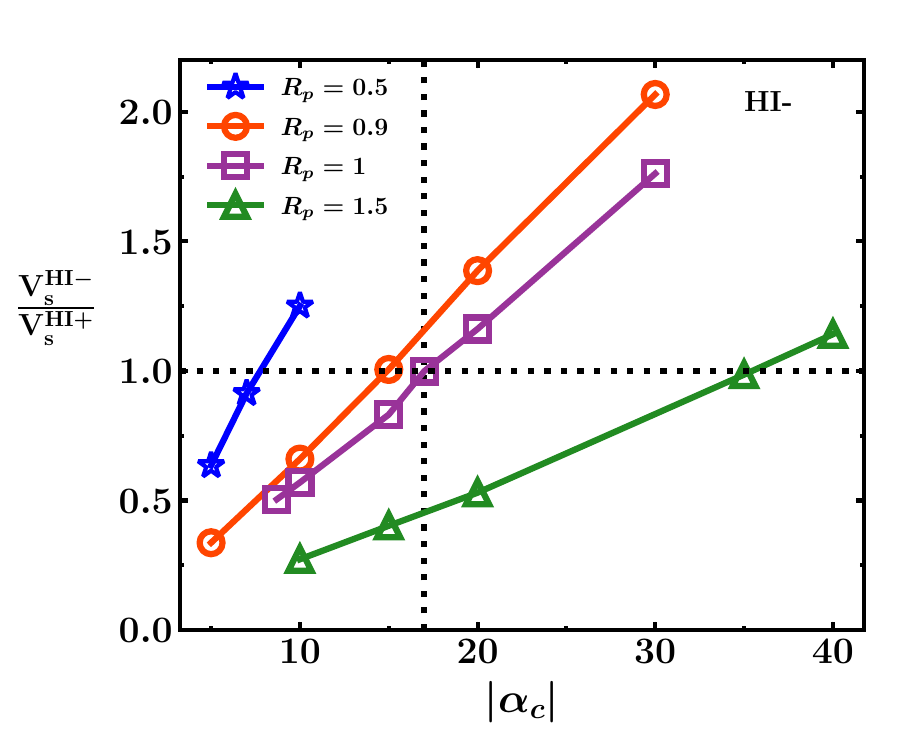}
     \caption{Self-propelled average velocity for dimers in the HI- case, normalized by the $V_s$ in HI+, for different values of $\alpha_{c}$ and particle radius. For each fixed particle radius $R_{p}$ there is a value of $\alpha_{c}$ that matches the HI radius, as indicated with a dotted line for the $R_p=1$. This optimal $|\alpha_{c}|$ value increases with $R_{p}$. }
     \label{fig:vzwithalpha}
 \end{figure}
 In principle, the value of $\mathbf{F}_{c,i}$ is determined by the product $\alpha_{c} {\bf\nabla} c({\bf r}_{i})$ which then also determines $V_s$. For clarity, we assume here that the concentration gradient ${\bf\nabla} c({\bf r}_{i})$ is given by Eq.~(\ref{eq:chemgrad}), with $\Delta c = \rho$, which reflects that the considered chemical reaction rate is $1$. 
With this and the value measured for the velocity with HI+ ($V_s=0.03$), we estimate $\alpha_c \simeq  8.8$. 
%, \mr{which already accounts also for the previous mismatch in the determination of the concentration.} \mrx{I am hesitant if keep this very last blue... } 
To check the validity of this estimation we measure the average dimer velocity in simulations with the \mbox{Ph-BD} model. The result in Fig.~\ref{fig:vzwithalpha} shows that for the above $\alpha_c$ value, the velocity is roughly a factor two smaller than the input value. We therefore perform simulations in the case of a single dimer without fluctuations and harmonic potential, and check that the velocity relation in Eq.~(\ref{eq:alphac}) is exactly reproduced. This implies that the above mismatch is related both, to the Brownian fluctuations, and to the fact that the harmonic spring results on the motion of the two beads and therefore $\alpha_c$ needs to be roughly doubled to get the desired value of $V_s$.  Further simulations with various values of $\alpha_c$ are also shown in Fig.~\ref{fig:vzwithalpha}, where it can be determined that $\alpha_c=17$ is the optimal value to match the MPCD velocity. %%
Other values of the particle radius are also simulated with Ph-BD and shown in Fig.~\ref{fig:vzwithalpha} normalized by the value obtained with MPCD.  The slope of the increase of $V_{s}$ with $|\alpha_{c}|$ decreases with for increasing  particle radius, and given a fixed value of $|\alpha_{c}|$, the velocity decreases with the radius. Relevant for us is that for a given particle radius $R_{p}$, there is always a value of $\alpha_{c}$ for which the HI+ and HI- cases result in the same velocity, which increases with increasing in $R_{p}$.

\begin{figure}
\centering
     \includegraphics[width=0.4\textwidth]{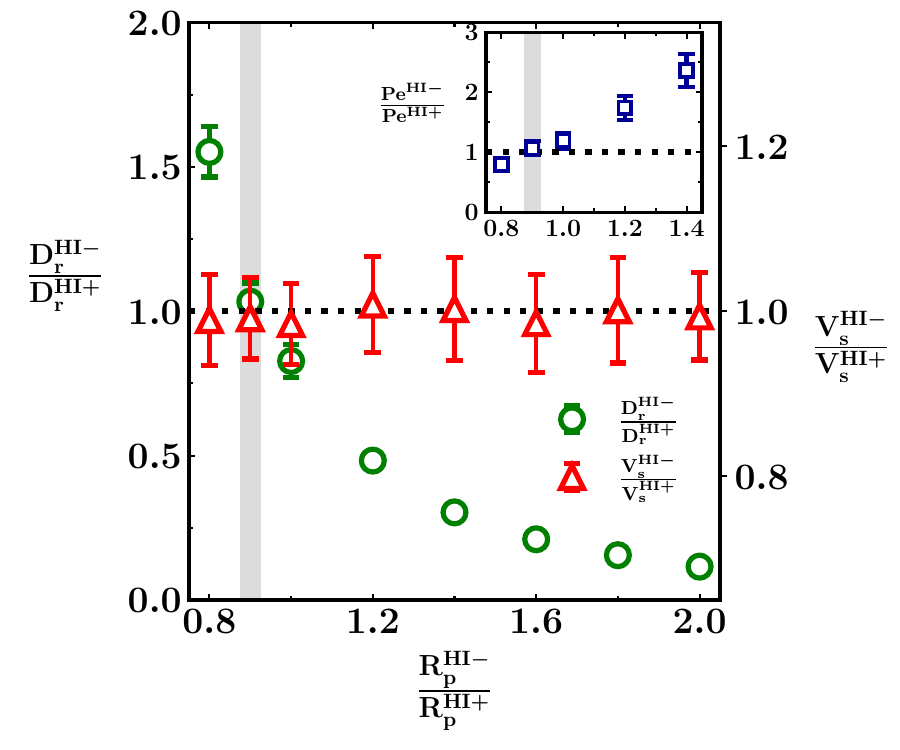}
     \caption{Normalized rotational diffusion and averaged velocity for Brownian dimers of different sizes with the $|\alpha_{c}|$ value that matches the HI velocity. Inset corresponding $Pe$ number, showing that this can also be matched by choosing a particular value for $R^{HI-}_{p}$. }
     \label{fig:vbd}
 \end{figure}
 The estimation in Eq.~(\ref{eq:dr}) for the rotational diffusion coefficient infers that in order to get the value $D_r=0.0072$ as obtained in the HI+ case, the effective particle radius should be $1.2$. However, the estimation is expected to be valid just for an spherical particle, and this is certainly different for the dimer here employed. We therefore explicitly measure $D_r$ in Ph-DB simulations  with different particle radius, $R_p^{HI-}$, which might differ from the value employed with MPCD. Simulations  with different $R_p^{HI-}$ use the value of $\alpha_c$ which optimizes $V_s$, as shown in Fig.~\ref{fig:vzwithalpha}. Results for $D_r$ are displayed in Fig.~\ref{fig:vbd} normalized by the MPCD measured value, together with the corresponding normalized $V_s$ values.
 While the results for $V_s$ in the HI- case are similar by construction to those of HI+, $D_r$ shows to decrease with increasing particle radius, which is the result of decrease in thermal noise for larger dimers. The corresponding P{\'e}clet numbers are calculated with Eq.~(\ref{eq:pec}) and the measured values of $V_s$, $D_r$, and are displayed in the inset of Fig.~\ref{fig:vbd}.

From these results, the particle radius $R_{p}^{HI-}=0.9 R_{p}^{HI+}$ emerges as optimal for mapping the two simulation methods, since $V_{s}$, as well as $D_r$ (and consequently also $Pe$) from both the models are in very good agreement with each other within the error of measurements. 
Therefore, for the dimer in HI- model we choose $|\alpha_{c}|=15$ with $R_p^{HI-}=0.9$, and $D_{r}=0.0075$, giving $Pe=4.4$, which is very close to $Pe=4.2$ in HI+ model.

 \begin{figure}[ht!]
    \centering
        \includegraphics[width=0.49\linewidth]{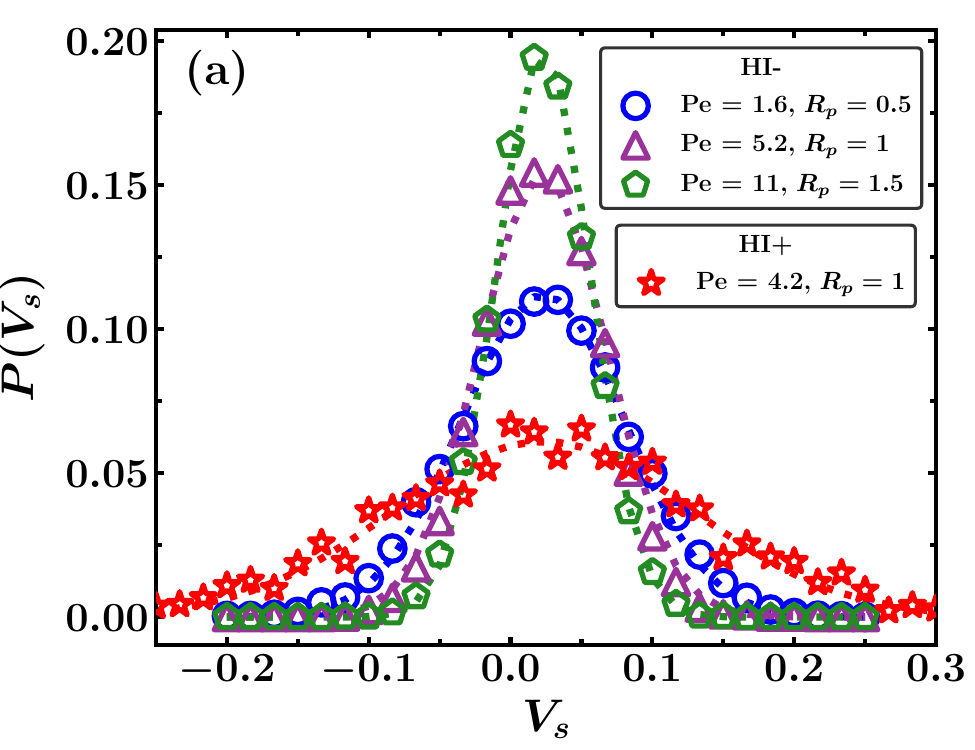}
        \includegraphics[width=0.49\linewidth]{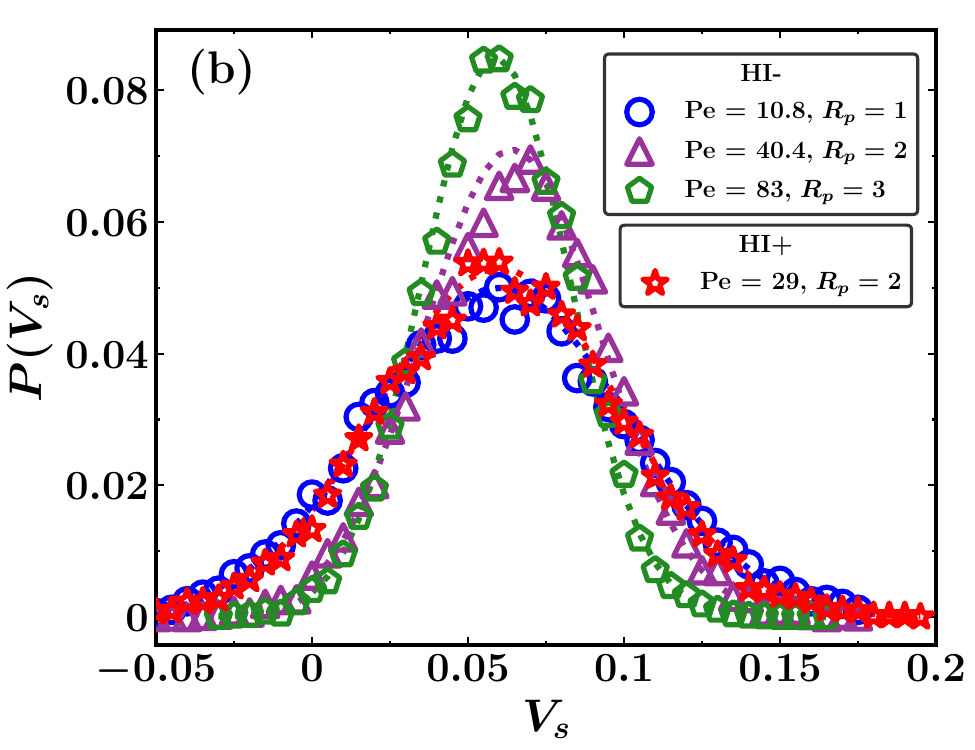}
    \caption{a)~Velocity probability distribution function for the HI+ case at $Pe=4.2$ and $R_{p}=1$, and for the HI- with matching average velocity and varying radius, which results in different $Pe$ numbers as indicated in the labels, showing that matching velocities have different distributions. b)~similar data for $R_{p}^{HI+}=2$, showing that the difference between distributions decreases with increasing particle radius.}
    \label{fig:veldis}
\end{figure}

Until now we have compared only averaged values, however the fluctuations, or similarly, which shape do the probability distributions of these quantities have, might also play a role. 
Therefore, in Fig.~\ref{fig:veldis} we quantify the distribution of $V_s$ for the HI+ and HI- cases, for various cases where the average velocity $\langle V_s \rangle$ is the same. Panel (a) shows results with $R_{p}^{HI+} =1$, while panel (b)  with $R_{p}^{HI+} =2$. Figure~\ref{fig:veldis}  shows that even though the average velocity is same, the distribution is always broader for the HI+ case, and that for HI- the distribution broadens for smaller bead radius. This can be understood since fluctuations have a larger effect for smaller particles in the presence of an explicit solvent.
Increasing the size of the dimer in the HI+ case shows to make the difference with HI- smaller. 
Nevertheless, we remain with $R_{p}^{HI+} = 1$, due to computational efficiency.

\section{\label{sec:polymer}Polar chemoattractive active polymer}

\begin{figure}[ht]
     \includegraphics[width=0.6\linewidth]{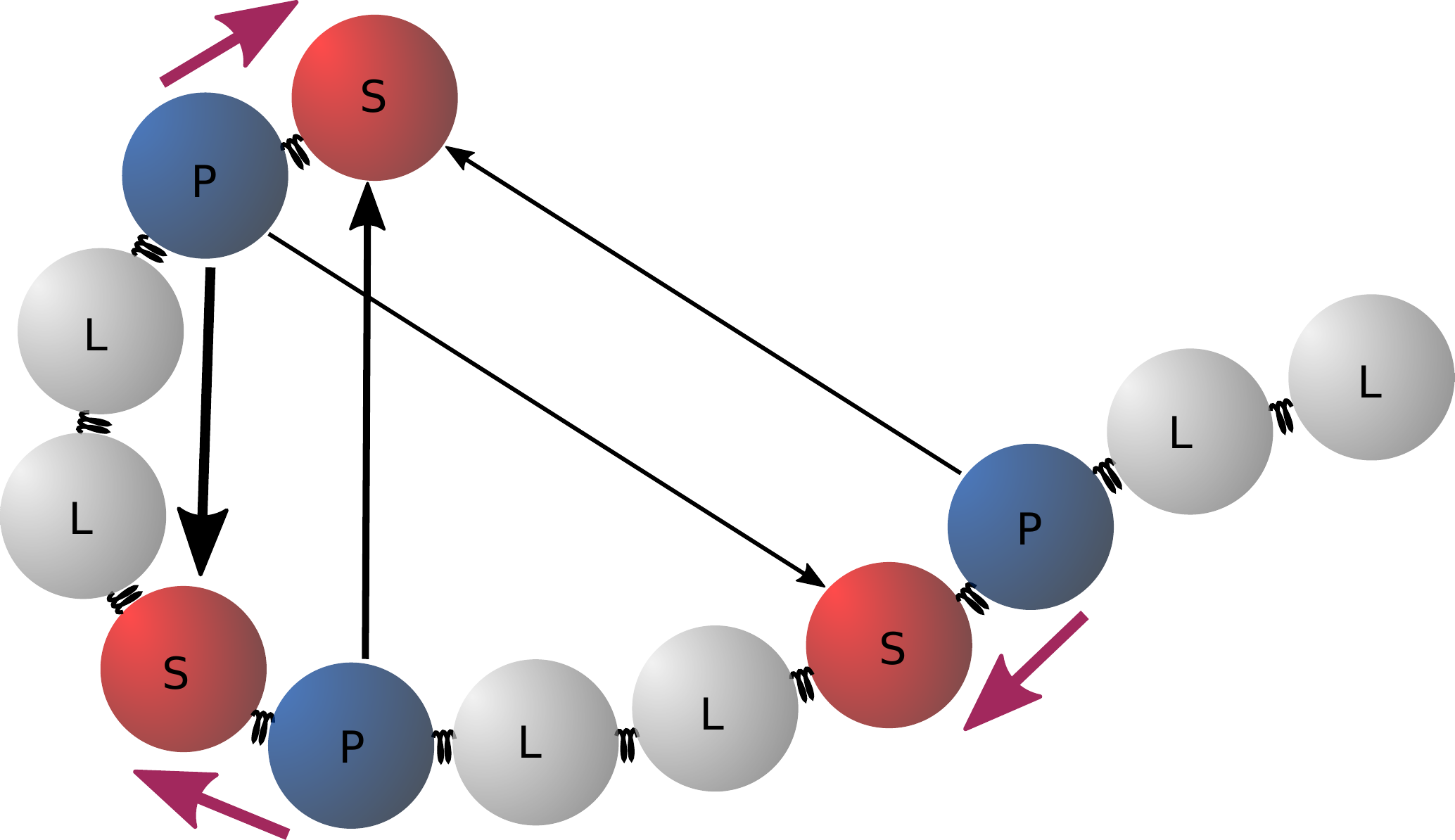}
     \caption{Polymer sketch with source beads ($S$) in red, phoretic beads ($P$) in blue and linker beads ($L$) in grey. Purple thick arrows indicate the direction of the local self-propelled force on the dimer, and black arrows indicate the phoretic attraction between distant source and phoretic beads.}
     \label{fig:schematic}
 \end{figure}

We employ now the described models and use the parameters optimized for the comparison with and without HI, and with these we investigate the properties first of a passive polymer and then of a chemoattractive active polymer.
The chemophoretic polymer is composed of a sequence of one source $S$, one phoretic $P$, and two linker $L$ beads in three-dimensions~\cite{namita2022collapse} (as shown in the sketch in Fig.~\ref{fig:schematic}) having $N$ monomers, namely $N/4$ sequences. As described in {\color{blue} Sec.~\ref{sec:phbd}}, $S$ accounts for beads with catalytic behavior, $P$ beads with a phoretic significant response, chosen here to be chemoattractive, and $L$ beads with no catalytic and no significant phoretic response.   
In this way, there is an effective local tangential force from each $P$ bead towards the $S$ direct neighbouring beads (purple arrows in Fig.~\ref{fig:schematic}), which provides the polymer with a well-defined polar active behavior. Furthermore, there is also long ranged attraction from the $P$ beads to all non-neighboring $S$ (black arrows in Fig.~\ref{fig:schematic}), which decreases with the beads separation, and provides the polymer with an additional attractive contribution. %%
In the HI+ case, these phoretic polymers have already shown to undergo a coil to globule transition by increasing the chemical activity~\cite{namita2022collapse}. However, those results were limited to relatively short polymers due to the high computational cost of the HI+ polymer model.  Here, we compare results of the HI+ and HI-, first to quantify the effect of HI, and then to extend the study in the HI- case to larger polymers. 
 
\subsection{Passive polymer
\label{sec:pol1}}

\begin{figure}[ht!]
 \centering
    \includegraphics[width=0.49\linewidth]{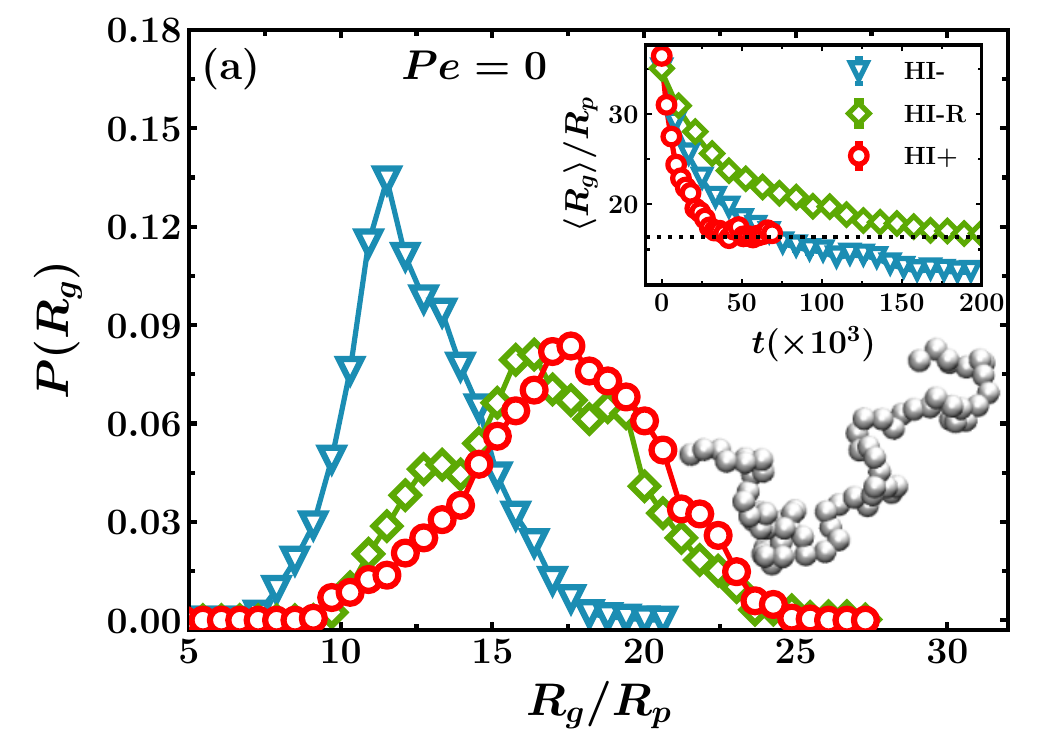}
    \includegraphics[width=0.49\linewidth]{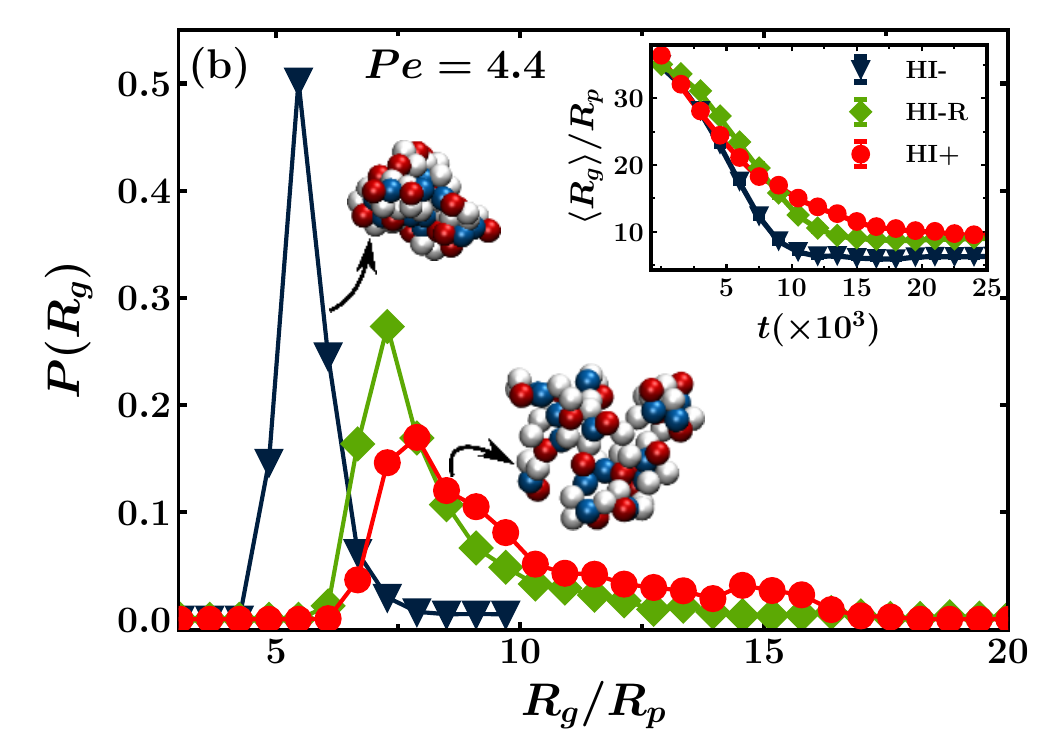}
    \caption{
    Radius of gyration probability distribution function for polymers with $N=76$ beads simulated with and without hydrodynamic interactions, HI+ and HI- respectively, and without HI but with additional soft repulsion HI-R. a)~Passive polymer, b)~chemoattractive active polymer.  %%
    Insets correspond to the time evolution of the radius of gyration averaged over ensembles, showing the typical relaxation time starting from a stretched configuration until reaching steady state in all cases. %%
    Included snapshots illustrate typical steady state configurations in a)~all passive cases are difficult to distinguish, so only the HI-R case is displayed, and in b)~the collapsed state with HI- is displayed together with the one from HI+, which clearly differ from each other. The HI-R case is very similar to HI+ and is not shown. See also related movie in Supplementary material.}
    \label{fig:prg}
\end{figure}

In order to test the proposed models, we first analyze the properties of a passive polymer  with and without HI by calculating the radius of gyration $R_g$, defined as
\begin{equation}\label{eq:radiusofgy}
    R_{g} = \sqrt{\frac{1}{N} \sum_{i=1}^{N}\left({\bf r}_{i}-{\bf R}_{cm}\right)^2},
\end{equation}
where ${\bf R}_{cm}$ is the center of mass of the chain and ${\bf r}_{i}$ is the position vector of $i^{th}$ bead. 
Since thermal fluctuations make that the polymer adopts a multitude of configurations with different values of  $R_g$, we calculate the probability distribution function $P(R_g)$ of polymers in the steady state. The results in Fig.~\ref{fig:prg}(a) for the passive polymer case show that the $R_g$ is displaced to larger values in the HI+ model with respect to the HI- model. 
%\emr{This means that the presence of explicit solvent induces an effective hydrodynamic repulsion between beads.} 
%{\color{blue} The origin of this repulsion could be due to an effective solvation layer present in the explicit solvent method MPCD, which is not present in an implicit solvent method like BD.} 
In principle, the presence of HI is not expected to affect the steady state average values of $R_g$ in equilibrium, as has previously checked in studies using computational methods with explicit~\cite{kikuchi2002polymer} and implicit solvents~\cite{Pham2008Browniandynamics}, respectively.
The origin of the mismatch observed in Fig.~\ref{fig:prg}(a) is therefore related to the combination of implicit (BD) and explicit solvent (MPCD) approaches here employed. The larger $R_g$ values obtained with MPCD can be understood as an effective repulsion given by a solvation layer which is present in experiments and also in computational methods with explicit solvent, but not in implicit solvent methods like BD.
In order to take this solvent effect into account in the BD simulations, we include a soft repulsive interaction as described for the RPh-BD model in {\color{blue} Sec.~\ref{sec:rphbd}}, referred here as HI-R. In Fig.~\ref{fig:prg}(a), the results for the distributions $P(R_g)$ using HI-R with $a_{ij}=25$ and $r^{\prime}_c=3$ show a very reasonable agreement with the distributions obtained with the HI+ model.  

Information about the characteristic dynamic behavior is obtained by quantifying the time evolution of the averaged radius of gyration for simulations starting from a stretched configuration.  
The time decay to the steady state value is shown in the inset of Fig.~\ref{fig:prg}(a). 
Previous works on passive polymers~\cite{kikuchi2002polymer,Pham2008Browniandynamics} have shown a faster collapse in presence of hydrodynamic interactions, which we clearly recover here. 
Curiously, the inclusion of repulsion in the Brownian model does not make the dynamics as faster as in the presence of HI, but the relaxation becomes even slower than for the purely Brownian case. This is an indication that the effective repulsion accounts for the additional solvent induced swelling of the polymer, but does not account for other solvent induced dynamic effects. 

\subsection{Active polymer: transition to a globule state}
\label{sec:pol2}
The radius of gyration is calculated also in simulations with activity, and $P(R_g)$,
the probability distribution function is shown for $Pe=4.4$ in Fig.~\ref{fig:prg}(b). 
Activity shows to induce in all cases a transition to a globule state, as clearly depicted by the shift of peak to lower $R_g$ values with respect to those in the passive case, similarly to previously described in the presence of HI~\cite{namita2022collapse}. This transition is due to the long range chemical gradient produced by the reaction on the $S$ bead.  The net chemical gradient then causes a subsequent effective attraction between $P$ and all other $S$ beads due to diffusiophoresis, thereby reducing the average size of the polymer.

While the transition to a globule state occurs both in the presence and absence of HI, 
the peak occurs at a lower $R_g$ for HI- model, which can be understood similarly as in the equilibrium case due to the presence of a certain solvent induced repulsion.  
The different compact globules are shown in the typical conformation snapshots in Fig.~\ref{fig:prg}(b).
The inclusion of the same repulsion force as in the passive case also prevents the polymer to collapse beyond a certain limit, such that there is only a small difference between the results with HI+ and HI-R. 
Similar to the passive case, the time evolution of the average $R_g$ with time in the presence of activity is shown in the inset of Fig.~\ref{fig:prg}(b). Activity makes that the relevant decay times are shorter than in the passive case. This is around 40\% faster in the HI+ case, and about  $5$ times faster in the HI- case. 
This implies that, opposite to the passive case, the presence of hydrodynamics (HI+) makes the polymer decay to its equilibrium size in a time almost twice as long than the purely Brownian case (HI-). Brownian with included repulsion (HI-R) polymers show a similar decay time as the HI+ case, which can not be directly explained as an extrapolation of the passive results. The effect of the hydrodynamic solvent has therefore non-trivial dynamic implications.            

\subsection{Structural properties dependence on polymer length} 
\label{sec:pol3}

\begin{figure}[ht!]
    \centering
        \includegraphics[width=0.49\linewidth]{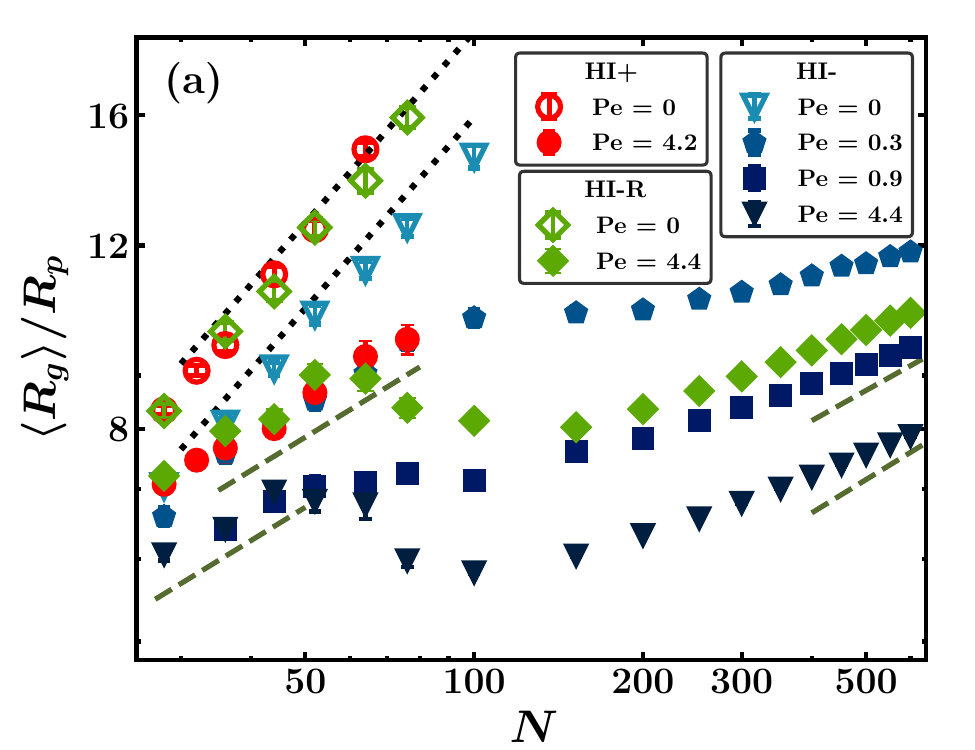}
        \includegraphics[width=0.49\linewidth]{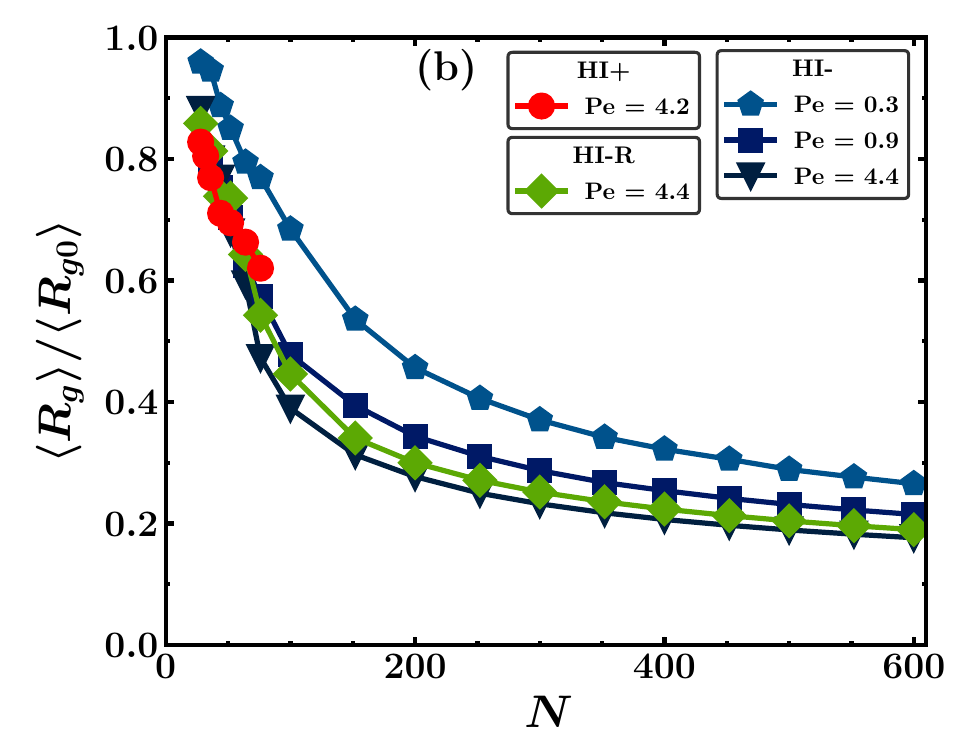}
    \caption{a)~Radius of gyration $R_{g}$ for polymers  of various sizes $N$ in steady state with and without HI, in equilibrium (empty symbols) and for different phoretic activities (full symbols), averaged over ensembles and time in the steady state. Note that the non-monotonous variation of $R_g$ with $N$ in the presence of activity. Dotted lines indicate a $N^{0.61}$ scaling for equilibrium configurations. Dashed lines are a guide to eye for polymers with activity in the intervals where $R_g$ grows with $N$, indicating $N^{0.33}$, the globular scaling, which is not yet reached for longer polymers. b)~Same data as in a) here normalized by the equilibrium values, showing a monotonous decrease of the relative radius of gyration with increasing activity and polymer size.}
    \label{fig:rgscaling}
\end{figure}
The radius of gyration normalized by the bead radius is depicted in Fig.~\ref{fig:rgscaling}(a) as a function of~$N$, for the passive case and for various activities. The effective polymer size given by~$R_g$ scales with  the number of monomers as~$R_{g} \sim N^{\nu}$, which determines the Flory exponent~$\nu$. 
For the passive polymer, the three models exhibit the expected scaling exponent $\nu_0~\approx 0.61$, indicating an open coil state of the polymer, as already reported for the HI+ case~\cite{namita2022collapse}, and illustrated in the snapshot in Fig.~\ref{fig:prg}(a). 

With the presence of activity the polymer collapses into a globule, as illustrated in the snapshots in Fig.~\ref{fig:prg}(b), which becomes smaller the larger the applied activity. 
For relatively short active chains, a scaling growth behavior with the number of monomers is also observed in Fig.~\ref{fig:rgscaling}, but with a smaller exponent than in the passive case, $\nu_a~\approx 0.33$, which is the signature of emergence of globular states. This exponent is the same for the three methods, namely it does not vary in the presence or absence of HI.  
The size of the polymers in HI+ shows a monotonous growth up to $N=76$ beads, and simulations with longer chains are not performed due to the large computational cost of including the explicit solvent.

Longer polymers are relatively easy to simulate in the absence of HI, such that we extend the investigation up to polymers of $N=600$ monomers. 
In the absence of HI, the  $R_g$  exhibits a non-monotonic behaviour. 
After the power law growth for short chains, 
for the intermediate polymer size ($50 \leq N \leq 100$) the $R_g$ decreases with $N$, which then starts rising again on increasing $N$. 
%{\color{blue}\sout{approaching the globular scaling of $\nu_a~\approx 0.33$.}} 
As evident from Fig.~\ref{fig:rgscaling}(a), the non-monotonic behaviour is activity dependent and becomes stronger with increasing activity. 
For an active chain, as we increase $N$, the number of active sites also increases thereby enhancing the net attraction between the polymeric segments. 
Simultaneously, increasing $N$ also leads to higher entropy of the system. 
The competition between the attractive active force and the entropy results into the non-monotonic feature in the $R_g$ scaling. 
The intermediate $N$ regime is where the attractive force dominates over entropy leading to more compact structures. 
For the HI+ case such non-monotonous behavior is not observed, however our data is not enough to conclude if this will simply occur at slightly larger $N$ values than for the HI- counterparts, or if the presence of HI is enough to compensate the effect of attraction. Simulations with the HI-R model, which are very similar to the HI+ case for short chains, also show clearly the non-monotonous growth of the $R_g$ with the polymer length. We therefore conclude that this non-monotonous growth is an intrinsic property of chemoattractive polymers at these low values of the activity. 
For long enough polymers, the radius of gyration grows again with increasing values of $N$. The growth exponents seem to vary with $Pe$ and the employed model, which will be studied elsewhere. The configurations show to be collapsed, and in Fig.~\ref{fig:rgscaling}(a) the globular scaling exponent $\nu_a={0.33}$ is displayed as a guide to eye.

To further understand the non-monotonous growth of $R_g$ with $N$, we display in Fig.~\ref{fig:rgscaling}(b) the $R_g$ of the chains with activity, normalized by their average radius of gyration in the passive case $\langle R_{g0}\rangle$. 
For short polymers, we employ the values explicitly calculated for each of the models in the passive case. For longer polymers, we extrapolate the values using the power law with the corresponding Flory exponent. The results indicate that the globules become progressively more compact with increasing $N$, and also with increasing $Pe$. This compression is faster for short polymer lengths and reaches a constant value for longer polymer lengths. This compression ratio is also almost the same for the three simulation models, showing that for this change in structure, HI or the soft repulsive interaction between monomers do not play a fundamental role in the formation of the compact globules.  

\subsection{ Dynamical Properties} 
\label{sec:pol4}

\begin{figure}[ht!]
    \centering
        \includegraphics[width=0.49\linewidth]{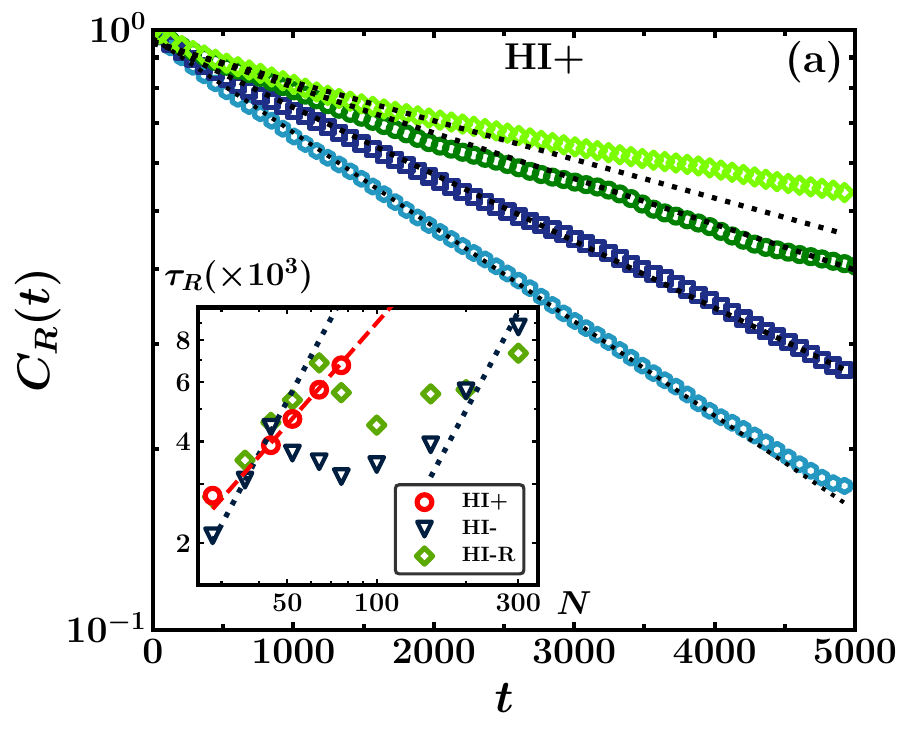}
        \includegraphics[width=0.49\linewidth]{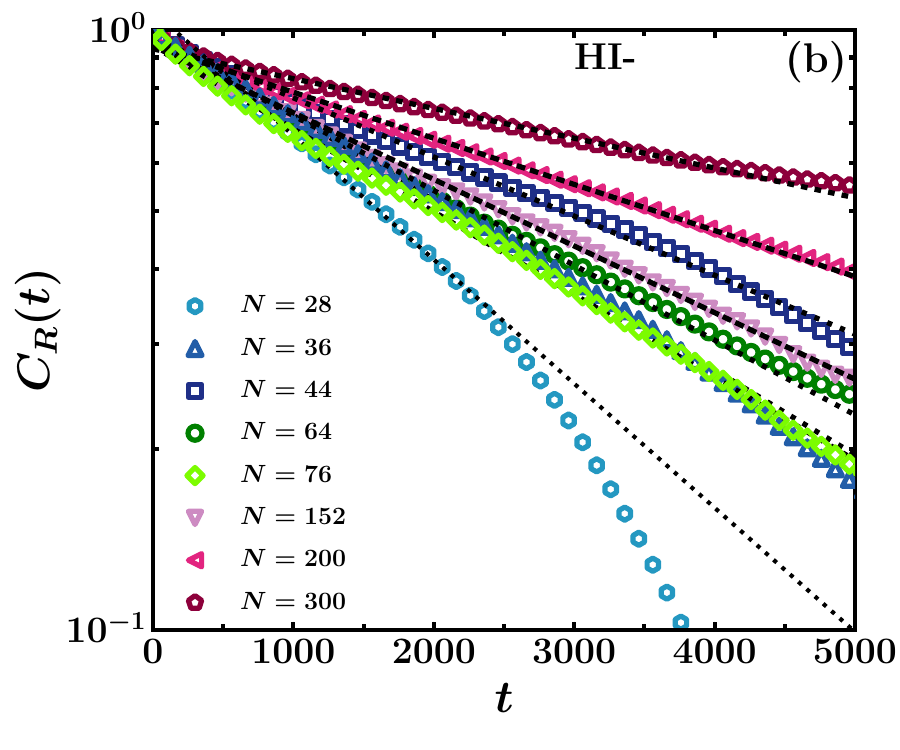}
    \caption{Time dependence of the end to end correlation function $C_R(t)$ for polymers of various lengths at $Pe=4.4$ a)~corresponds to HI+ results, and b)~to HI-. Symbols are calculated from the simulations with Eq.~(\ref{corr}) for various polymer lengths. Discontinuous lines are the fits to the exponential $C_R(t)\sim e^{-t/\tau_R}$.  In the inset of a) the obtained characteristic relaxation times $\tau_R$ are displayed as a function of the polymer length $N$. The dotted lines in the inset indicate the $N^{1.66}$ scaling as expected from the Rouse dynamics for short polymer lengths. The dashed lines correspond to the scaling as $N$, confirming the Zimm dynamics.
The non-monotonous behaviour for longer chains show similar dependence to Fig.~\ref{fig:rgscaling}(a).}
    \label{fig:corr}
\end{figure}
The polymer dynamical properties have been shown to be affected by activity either by enhancing its directional velocity, in the case of semi-flexible polymers~\cite{debarati2016solutegradient}, or by amplifying its effective diffusion, in case of flexible polymers~\cite{biswas2017catalystcoated}. Here, we probe the relaxation dynamics of the polymer by calculating the time correlation $C_R(t)$ of the normalized end-to-end vector ${\bf \widehat{R}}_{E}(t)=[{\bf r}_{N}(t)-{\bf r}_{1}(t)]/|{\bf r}_{N}(t)-{\bf r}_{1}(t)|$ defined as
\begin{equation}\label{corr}
    C_R(t) = \big\langle {\bf \widehat {R}}_{E}(t-t_{0}) \cdot {\bf \widehat {R}}_{E}(t_{0})\big\rangle, 
\end{equation}
and calculated in the steady state, namely in time intervals where there is no dependence on the initial time $t_{0}$.
We plot the $C_R(t)$ for HI+ and HI- models in Fig.~\ref{fig:corr}(a) and Fig.~\ref{fig:corr}(b), respectively. 
The decay of the correlation function very reasonably agrees with an exponential decay 
$C_R(t)\sim \exp(-t/\tau_R)$~\cite{doi1988theory}, for a large range of $N$ values.
The characteristic relaxation time of the polymer $\tau_R$ can then be calculated by fitting the exponential decay of $C_R(t)$. Results are shown in the inset of Fig.~\ref{fig:corr}(a). 

For HI+ model in Fig.~\ref{fig:corr}(a), the correlation function decays slower with time the larger are the polymers, this is there is a systematic increase in $\tau_R$, as also shown in the inset.
For HI- model in Fig.~\ref{fig:corr}(b), it can be seen that the correlation function decays with time does not show a monotonous behavior with increasing polymer length. This trend is easier observed in the dependence of $\tau_R$ with $N$ in the inset of Fig.~\ref{fig:corr}(a) where the data obtained for the HI-R shows the same dependence, with slightly larger values of $\tau_R$. These larger times are in agreement with the non-steady state relaxation times discussed in {\color{blue}Sec.~\ref{sec:pol2}}, and shown in the inset of Fig.~\ref{fig:prg}(b). For short polymer lengths, the values of $\tau_R$ scale with $N^{3\nu_a}$ for the HI+ case (this is $N$), and with $N^{1+2\nu_a}$ for both HI- and HI-R (this is $N^{1.66}$), in agreement with Zimm and Rouse dynamics~\cite{doi1988theory}, respectively. For medium size polymers without HI, a non-monotonous decrease of $\tau_R$ with $N$ behavior is observed, similar to the radius of gyration. 
%\emr{For long polymers, $\tau_R$ increases again with $N$, with a scaling that seems consistent with Rouse in the HI- case, but it is not so clear in the HI-R case.}   
The non-monotonous behavior could be due to the enhancement in the effective attraction between the monomers at these  intermediate $N$ values which lead to smaller values of $R_g$  and also to faster movement of the polymer and hence lower $\tau_R$. For long polymers, $\tau_R$ increases again with $N$, with a scaling that seems consistent with Rouse in the HI- case, but it is not so clear in the HI-R case. The difference between the results of these two methods has the additional value of illustrating the relevance of the solvation layer, which is expected to vary considering specific details of different systems with experimental relevance.

Hence, both the structural and dynamical quantification indicates that the presence of the competing attractive, steric and entropic forces in the polymer leads to a non-monotonic behaviour. We therefore expect that this behaviour will also be seen in the presence of HI for longer polymers.  

\section{Summary and conclusions}
\label{sec:conclusion}

An effective implicit method to simulate the properties of chemophoretic colloidal based systems is here proposed and optimized for the case of chemoattractive polymers, in the absence of HI. 
While in principle phoresis is  a solvent induced phenomena, this method treats the phoretic thrust as an effective interaction between colloids, or monomer beads. 
Similar to experimental conditions, the phoretic thrust has the same nature when it translates into self-propulsion, or interparticle induced repulsion, attraction, or torque.  
The benefits of the proposed method are twofold. On the one hand, the method parameters  can be chosen to closely map another method where hydrodynamic interactions are explicitly considered, such that a direct comparison of the results obtained with the two methods enables the quantification of the effect of HI in the relevant system properties.
On the other hand, the absence of explicit solvent makes it easier to investigate the phoretic properties of much larger systems.

For polymers in equilibrium conditions, the presence of explicit solvent results into an additional soft repulsion among beads, together with an enhancement of the typical relaxation time.
The phoretic Brownian dynamics model is then extended to consider the effect of an additional soft repulsion, which mimics the effect of a solvation layer effect also present in experimental systems. 
For the chemoattractive active polymers here investigated, a transition to a globular state is observed in all models. The repulsive Brownian model accurately recovers configurations of the active hydrodynamic model, which are more compact in the BD model, due to absence of solvent, similar to the passive case.
However, opposite to the passive case, in the presence of activity the dynamic relaxation time gets faster when hydrodynamic interactions are considered. This indicates first that the the effect of hydrodynamics can be non-trivial, and second that repulsive Brownian dynamics method is a very promising tool that can be used to precisely evaluate their effect. 
Long chemoattractive polymers show a relative shrinkage, which increases both with polymer size and with activity. 
However, the absolute change of size with polymer size, and of the characteristic relaxation time of the end-to-end relaxation time is non-monotonous, which is due to the balance of two opposite effects. 
The larger number of active phoretic monomers as the polymer size increases makes the overall attraction between monomers  stronger, facilitating the collapse into more compact structures.  
Simultaneously, the increase of entropy, due to the increase of local tangential force, facilitates the polymer extension.  
In conclusion, the phoretic Brownian dynamic models here proposed are shown to constitute a very competitive alternative for the simulation of diffusiophoretic processes, which we expect will trigger a number of related studies of phoretic synthetic materials, or even in the correlated motion of active units on chromosomes where similar mechanisms play an important role.

\begin{acknowledgments}
\noindent
S. J. thanks a research fellowship from UGC India.
S. T. acknowledges SERB India for research grant number CRG/2022/003778.
The authors would like to acknowledge the HPC facility at IISER Bhopal for allocating the necessary computational resources. 
\end{acknowledgments}

\section*{References}
%\bibliography{ref}
%\bibliographystyle{unsrt}

\end{document}